\documentclass[fleqn,twoside,twocolumn,nofootinbib,showkeys]{revtex4} 
\usepackage[sec,nocpr]{ujp} 

\begin{document}
\title[Lyapunov Indices and the Poincar\'{e} Mapping]
{LYAPUNOV INDICES AND THE POINCAR\'{E} MAPPING IN A STUDY OF THE
STABILITY OF THE KREBS CYCLE}%
\author{V.I. Grytsay}
\affiliation{\bitp}
\address{\bitpaddr}
\email{vgrytsay@bitp.kiev.ua}

\udk{577.3} \pacs{05.45.-a, 05.45.Pq,\\[-3pt] 05.65.+b} \razd{\secx}

\autorcol{V.I.\hspace*{0.7mm}Grytsay}

\setcounter{page}{561}%

\begin{abstract}
On the basis of a mathematical model, we continue the study of the
metabolic Krebs cycle (or the tricarboxylic acid cycle).\,\,For the
first time, we consider its consistency and stability, which depend
on the dissipation of a transmembrane potential formed by the
respiratory chain in the plasmatic membrane of a cell.\,\,The
phase-parametric characteristic of the dynamics of the ATP level
depending on a given parameter is constructed.\,\,The scenario of
formation of multiple autoperiodic and chaotic modes is presented.
Poincar\'{e} sections and mappings are constructed.\,\,The stability
of modes and the fractality of the obtained bifurcations are
studied.\,\,The full spectra of Lyapunov indices, divergences,
KS-entropies, horizons of predictability, and Lyapunov
dimensionalities of strange attractors are calculated.\,\,Some
conclusions about the structural-functional connections determining
the dependence of the cell respiration cyclicity on the
synchronization of the functioning of the tricarboxylic acid cycle
and the electron transport chain are presented.
\end{abstract}
 \maketitle

\noindent The most important task of synergetics is the search for
the general physical laws explaining the natural regularity of the
formation of a life on the Earth.\,\,The first descriptive
experiment demonstrating the possibility for a cyclic metabolic
process to exist was executed by B.P.~Belousov in 1951 [1].\,\,With
the help of some chemical substances (citric acid, potassium
bromate, cerium, {\it etc.}), he successfully constructed a model of
autooscillatory metabolic process involving the Krebs cycle
[2].\,\,By this, he showed for the first time that the vital
metabolic processes in a cell can be supported due to their
self-organization in the autoperiodic mode.

The tricarboxylic acid cycle occupies a particular place in the
vital activity of aerobic cells.\,\,As a result of the cyclic
metabolic process, the acetyl groups formed in the decay of
carbohydrates, fats, and proteins are oxidized there to carbon
dioxide.\,\,Hydrogen atoms released in this case are transferred
into the respiratory chain, where the main source of the energy for
a cell, ATP, is produced in the course of the reaction of oxidative
phosphorylation.\,\,With the help of NADH, there arise the negative
feedbacks, due to which the synchronization of the process of
catabolism and the respiration of a cell occurs.\,\,The joint
existence of these metabolic processes is possible only under their
self-organization in a single cycle.\,\,In addition, the Krebs cycle
is a source of molecules-precursors, which are used in the synthesis
of compounds important for the vital activity of cells in other
biochemical reactions.

Studying the functioning of the tricarboxylic acid cycle was carried
out in experiments and theoretically [3--11].\,\,In particular, the
given process was analyzed on the basis of the mathematical model of
growth of \textit{Candida utilis} cells on ethanol.\,\,This model
was developed by Professor V.P.~Gachok [12, 13].\,\,The analogous
approaches to the modeling of a growth of cells were considered by
J.\,\,Monod, V.S.\,\,Pod\-gors\-kii, L.N.\,\,Drozdov-Tikhomirov,
N.T.\,\,Rakhimova, G.Yu.\,\,Riz\-ni\-chen\-ko, and others
[14--18].\,\,Within such models, the unstable modes in the
cultivation of cells, which were observed in experiments, were
stu\-died.\,\,The results of computational experiments concerning
the chaotic dynamics described well the experimental characteristics
[19].

Then the Gachok model was improved in [20, 21], where the influence
of the concentration of CO$_{2}$ on the cell respiration intensity
and the cyclicity of a respiratory process was taken into
account.\,\,The struc\-tu\-ral-functional connections of the
metabolic process running in a cell, according to which the
appearance of complicated oscillations in the metabolic process in a
cell becomes possible, were found.\,\,It was concluded that those
oscillations arise at the level of redox reactions of the Krebs
cycle, reflect the cyclicity of the process, and characterize the
self-organization inside a cell.\,\,For some modes, the fractality
of bifurcations was studied, and the indicators characterizing the
stability of strange attractors were established.

Analogous oscillatory modes in the processes of photosynthesis and
glycolysis, variations in the concentration of calcium in a cell,
and oscillations in a heart muscle and in some biochemical processes
were found in [22--26].

The distinction of the present work from other ones consists in the
modeling of such significant phenomenon as the influence of the
dissipation of a proton potential on the Krebs cycle.\,\,By the
Mitchell chemoosmotic hypothesis [27], the transmembrane potential
arises under the reducing equivalent transfer along the respiratory
chain on internal mitochondrial memb\-ra\-nes.\,\,Then the transport
of ions H$^+$ inward a membrane occurs under the action of the
formed electrochemical gradient of the potential.\,\,This results in
the production of a free energy, whose significant part is stored in
ATP.\,\,A part of this energy is used also for other purposes,
namely: transport of phosphate, transport of ions Ca$^{2+}$,
transformation of ADP to ATP, generation of heat, operation of the
``proton-driven turbine'' of bacterial flagella, {\it etc.} The
transfer of ions H$^+$ through ${\rm F}_0 {\rm F}_1 -{\rm ATP}$ase
molecules changes the given potential, which affects the
tricarboxylic acid cycle.\,\,An analogous process is running in the
plasmatic membrane of aerobic cells \textit{Candida utilis}
considered in the present work.\,\,Thus, the Krebs cycle and the
respiratory chain are functioning under a permanent
self-organization between each other and depend on the intensity of
a dissipation of the electrochemical gradient of the potential in
metabolic processes.

We will study the structural-functional connections, according to
which the Krebs cycle and the respiratory chain are self-organized
and operate as a single complex providing a cell with the necessary
energy store for its life.\,\,The limits of stability of the cycle
depending on the dissipation of the proton potential in various
processes of the vital activity of a cell are considered as well.

\section{Mathematical Model}

The general scheme of the process is presented in
Fig.~1.\,\,According to it with regard for the mass balance, we have
constructed the mathematical model given by Eqs.~(1)--(19).
\[
\frac{dS}{dt}=S_0 \frac{K}{K+S+\gamma \psi }\,-
\]\vspace*{-7mm}
\begin{equation}
\label{eq1} -\,k_1 V(E_1 )\frac{N}{K_1 +N}V(S)-\alpha _1 S,
\end{equation}\vspace*{-7mm}
\[
\frac{dS_1 }{dt}=k_1 V(E_1 )\frac{N}{K_1 +N}V(S)\,-
\]\vspace*{-7mm}
\begin{equation}
\label{eq2} -\,k_2 V(E_2 )\frac{N}{K_1 +N}V(S_1 ),
\end{equation}\vspace*{-7mm}
\[
\frac{dS_2 }{dt}=k_2 V(E_2 )\frac{N}{K_1 +N}V(S_1 )\,-
\]\vspace*{-8mm}
\begin{equation}
\label{eq3} -\,k_3 V(S_2^2 )V(S_3 )-k_4 V(S_2 )V(S_8 ),
\end{equation}\vspace*{-7mm}
\[
\frac{dS_3 }{dt}=k_4 V(S_2 )V(S_8 )-k_5 V(N^2)V(S_3^2 )\,-
\]\vspace*{-8mm}
\begin{equation}
\label{eq4} -\,k_3 V(S_2^2 )V(S_3 ),
\end{equation}\vspace*{-7mm}
\[
\frac{dS_4 }{dt}=k_5 V(N^2)V(S_3^2 )-k_7 V(N)V(S_4 )\,-
\]\vspace*{-8mm}
\begin{equation}
\label{eq5} -\,k_8 V(N)V(S_4 ),
\end{equation}\vspace*{-7mm}
\begin{equation}
\label{eq6} \frac{dS_5 }{dt}=k_7 V(N)V(S_4 )-2k_9 V(L_1 -T)V(S_5 ),
\end{equation}\vspace*{-7mm}
\[
\frac{dS_6 }{dt}=2k_9 V(L_1 -T)V(S_5 )\,-
\]\vspace*{-8mm}
\begin{equation}
\label{eq7} -\,k_{10} V(N)\frac{S_6^2 }{S_6^2 +1+M_1 S_8 },
\end{equation}\vspace*{-7mm}
\[
\frac{dS_7 }{dt}=k_{10} V(N)\frac{S_6^2 }{S_6^2 +1+M_1 S_8 }-k_{11}
V(N)V(S_7 )\,-
\]\vspace*{-7mm}
\begin{equation}
\label{eq8} -\,k_{12} \frac{S_7^2 }{S_7^2 +1+M_2 S_9 }V(\psi ^2)+k_3
V(S_2^2 )V(S_3 ),
\end{equation}\vspace*{-7mm}
\[
\frac{dS_8 }{dt}=k_{11} V(N)V(S_7 )-k_4 V(S_2 )V(S_8 )\,+
\]\vspace*{-7mm}
\begin{equation}
\label{eq9} +\,k_6 V(T^2)\frac{S^2}{S^2+\beta _1 } \frac{N_1 }{N_1
+(S_5 +S_7 )^2},
\end{equation}\vspace*{-7mm}
\[
\frac{dS_9 }{dt}=k_{12} \frac{S_7^2 }{S_7^2 +1+M_2 S_9 }V(\psi
^2)\,-
\]\vspace*{-7mm}
\begin{equation}
\label{eq10} -\,k_{14} \frac{XTS_9 }{(\mu _1 +T)[(\mu _2 +S_9 +X+M_3
(1+\mu _3 \psi )]S},
\end{equation}\vspace*{-7mm}
\[
\frac{dX}{dt}=k_{14} \frac{XTS_9 }{(\mu _1\! +T)[(\mu _2\! +S_9\!
+X+M_3 (1+\mu _3 \psi )]S}\,-
\]\vspace*{-7mm}
\begin{equation}
\label{eq11} -\,\alpha _2 X,
\end{equation}\vspace*{-7mm}
\[
\frac{dQ}{dt}=-k_{15} V(Q)V(L_2 -N)\,+
\]\vspace*{-7mm}
\begin{equation}
\label{eq12} +\,4k_{16} V(L_3 -Q)V(O_2 )\frac{1}{1+\gamma _1 \psi
^2},
\end{equation}\vspace*{-7mm}
\[
\frac{dO_2 }{dt}=O_{2_0 } \frac{K_2 }{K_2 +O_2 }-k_{16} (L_3
-Q)V(O_2 )\frac{1}{1+\gamma _1 \psi }\,-
\]\vspace*{-7mm}
\begin{equation}
\label{eq13} -\,k_8 V(N)V(S_4 )-\alpha _3 O_2,
\end{equation}\vspace*{-7mm}
\[
\frac{dN}{dt}=-k_7 V(N)V(S_4 )-k_{10} V(N)\frac{S_6^2 }{S_6^2 +1+M_1
S_8 }\,-
\]\vspace*{-7mm}
\[
-\,k_{11} V(N)V(S_7 )-k_5 V(N^2)V(S_3^2 )\,+
\]\vspace*{-7mm}
\[
+\,k_{15} V(Q)V(L_2 -N)-k_2 V(E_2 )\frac{N}{K_1 +N}V(S_1 )\,-
\]\vspace*{-7mm}
\begin{equation}
\label{eq14} -\,k_1 V(E_1 )\frac{N}{K_1 +N}V(S),
\end{equation}\vspace*{-7mm}
\[
\frac{dT}{dt}=k_{17} V(L_1 -T)V(\psi ^2)+k_9 V(L-T)V(S_3 )-\alpha _4
T\,-
\]\vspace*{-7mm}
\[
 -\,k_{18} k_6 V(T^2)\frac{S^2}{S^2+\beta _1 }
\frac{N_1 }{N_1 +(S_5 +S_7 )^2}\,-
\]\vspace*{-7mm}
\begin{equation}
\label{eq15}-\,k_{19} k_{14} \frac{XTS_9 }{(\mu _1\! +T)[\mu _2\!
+S_9\! +\!X\!+M_3 (1+\mu _3 \psi )S]},
\end{equation}\vspace*{-7mm}
\[
\frac{d\psi }{dt}=4k_{15} V(Q)V(L_2 -N)+4k_{17} V((L_1 -T)V(\psi
^2)\,-
\]\vspace*{-7mm}
\begin{equation}
\label{eq16} -\,2k_{12} \frac{S_7^2 }{S_7^2 +1+M_2 S_9 }V(\psi
^2)-\alpha \psi,
\end{equation}\vspace*{-7mm}
\[
\frac{dE_1 }{dt}=E_{1_0 } \frac{S^2}{\beta _2 +S^2}\frac{N_2 }{N_2
+S_1 }\,-
\]\vspace*{-7mm}
\begin{equation}
\label{eq17} -\,n_1 V(E_1 )\frac{N}{K_1 +N}V(S)-\alpha _5 E_1,
\end{equation}\vspace*{-7mm}
\[
\frac{dE_2 }{dt}=E_{2_0 } \frac{S_1^2 }{\beta _3 +S_1^2 }\frac{N_3
}{N_3 +S_2 }\,-
\]\vspace*{-7mm}
\begin{equation}
\label{eq18} -\,n_2 V(E_2 )\frac{N}{K_1 +N}V(S_1 )-\alpha _6 E_2,
\end{equation}\vspace*{-7mm}
\begin{equation}
\label{eq19} \frac{dC}{dt}=k_8 V(N)V(S_4 )-\alpha _7 C,
\end{equation}
where $V(X)=X/(1+X)$ is the function that describes the adsorption
of the enzyme in the region of a local coupling.\,\,The variables of
the system are dimensionless [12, 13].

The internal parameters of the system are as follows:
\[
k_1 =0.3; \quad k_2 =0.3; \quad k_3 =0.2; \quad k_4 =0.6;
\]\vspace*{-9mm}
\[
 k_5
=0.16; \quad k_6 =0.7; \quad k_7 =0.08; \quad k_8 =0.022;
\]\vspace*{-9mm}
\[
k_9 =0.1; \quad k_{10} =0.08; \quad k_{11} =0.08; \quad k_{12} =0.1;
\]\vspace*{-9mm}
\[
k_{14} =0.7; \quad k_{15} =0.27; \quad k_{16} =0.18;
\]\vspace*{-9mm}
\[
k_{17} =0.14; \quad k_{18} =1; \quad k_{19} =10; \quad n_1 =0.07;
\]\vspace*{-9mm}
\[
 n_2 =0.07; \quad L=2; \quad L_1 =2; \quad L_2 =2.5; \quad L_3
=2;
\]\vspace*{-9mm}
\[
K=2.5; \quad K_1 =0.35; \quad K_2 =2; \quad M_1 =1;
\]\vspace*{-9mm}
\[
M_2 =0.35; \quad M_3 =1; \quad N_1 =0.6; \quad N_2 =0.03;
\]\vspace*{-9mm}
\[
N_3 =0.01; \quad \mu _1 =1.37; \quad \mu _2 =0.3; \quad \mu _3
=0.01;
\]\vspace*{-9mm}
\[
\gamma =0.7; \quad \gamma _1 =0.7; \quad \beta _1 =0.5; \quad \beta
_2 =0.4;
\]\vspace*{-9mm}
\[
\beta _3 =0.4; \quad E_{1_0 } =2; \quad E_{2_0 } =2.
\]
The external parameters determining the flow-type conditions are
chosen as
\[
S_0 =0.05055; \quad O_{2_0 } =0.06; \quad \alpha =0.002;
\]\vspace*{-9mm}
\[
 \alpha _1
=0.02; \quad \alpha _2 =0.004; \quad \alpha _3 =0.01;
\]\vspace*{-9mm}
\[
\alpha _4 =0.01; \quad \alpha _5 =0.01; \quad \alpha _6 =0.01; \quad
\alpha _7 =0.0001.
\]
The model covers the processes of substrate-en\-zy\-ma\-tic
oxidation of ethanol to acetate, the cycle involving tri- and
dicarboxylic acids, glyoxylate cycle, and respiratory chain.

\begin{figure*}%
\vskip1mm
\includegraphics[width=12cm]{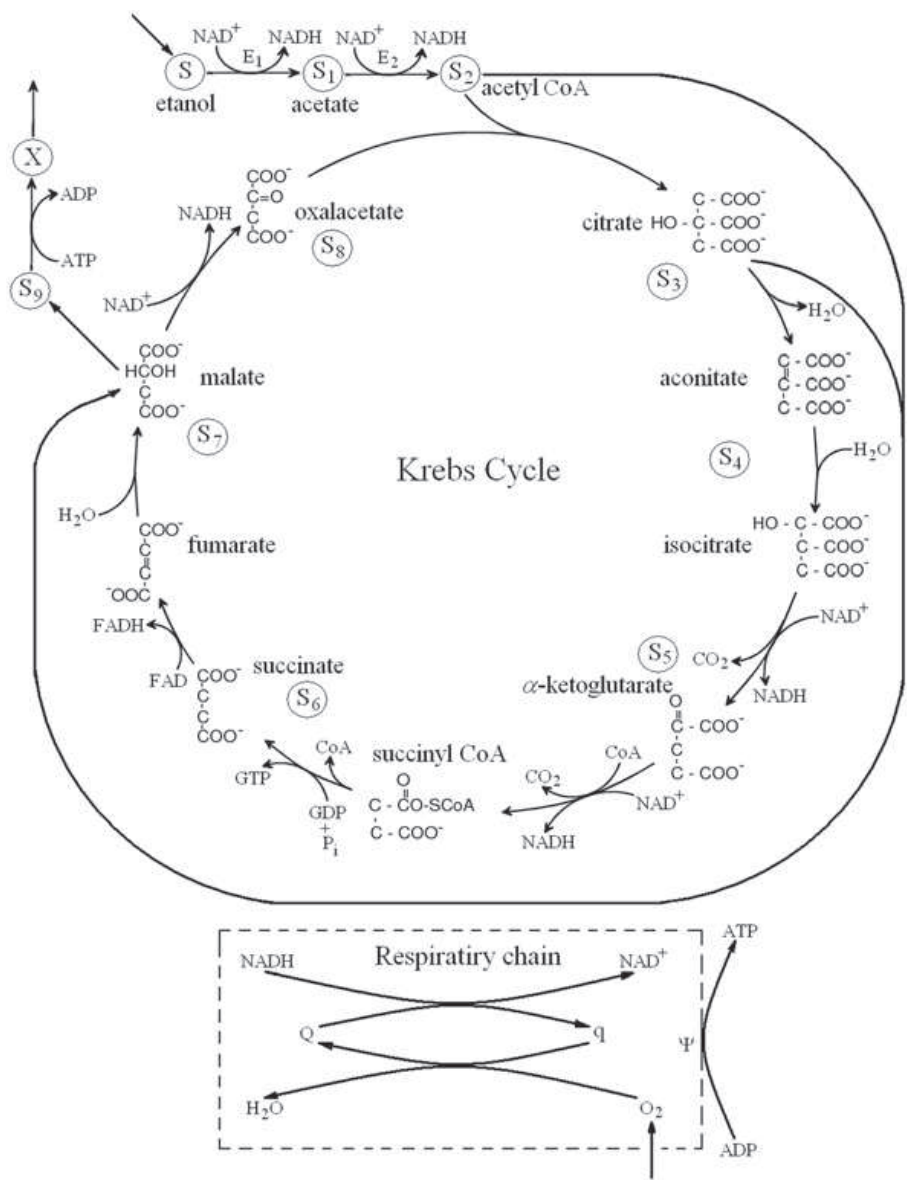}
\vskip-2mm\parbox{12cm}{\caption{General scheme of the metabolic
process of growth of cells \textit{Candida utilis} on ethanol  }}
\end{figure*}

The incoming ethanol $S$ is oxidized by the alcohol dehydrogenase
enzyme $E_1 $ to acetaldehyde $S_1 $ (\ref{eq1}) and then by the
acetal dehydrogenase enzyme $E_2 $ to acetate $S_2 $ (\ref{eq2}),
(\ref{eq3}). The formed acetate can participate in the cell
metabolism and can be exchanged with the environment. The model
accounts for this situation by the change of acetate by acetyl-CoA.
On the first stage of the Krebs cycle due to the citrate synthase
reaction, acetyl-CoA jointly with oxalacetate $S_8 $ formed in the
Krebs cycle create citrate $S_3 $ (\ref{eq4}). Then substances $S_4
-S_8 $ are created successively on stages (\ref{eq5})--(\ref{eq9}).
In the model, the Krebs cycle is represented by only those
substrates that participate in the reduction of NADH and the
phosphorylation ADT~$\to $~ATP. Acetyl-CoA passes along the chain to
malate represented in the model as intramitochondrial $S_7 $
(\ref{eq8}) and cytosolic $S_9 $ (\ref{eq10}) ones. Malate can be
also synthesized in another way related to the activity of two
enzymes: isocitrate lyase and malate synthetase.\,\,The former
catalyzes the splitting of isocitrate to succinate, and the latter
catalyzes the condensation of acetyl-CoA with glyoxylate and the
formation of malate.\,\,This glyoxylate-linked way is shown in
Fig.~1 as an enzymatic reaction with the consumption of $S_2 $ and
$S_3 $ and the formation of $S_7 $.\,\,The parameter $k_3 $ controls
the activity of the glyoxylate-linked way (\ref{eq3}), (\ref{eq4}),
(\ref{eq8}). The yield of $S_7 $ into cytosol is controlled by its
concentration, which can increase due to $S_9$, by causing the
inhibition of its transport with the participation of protons of the
mitochondrial
membrane.

The formed malate $S_9 $ is used by a cell for its growth, namely
for the biosynthesis of protein $X$ (\ref{eq11}). The energy
consumption of the given process is supported by the process
ATP~$\to $~ADP.\,\,The presence of ethanol in the external solution
causes the ``ageing'' of external membranes of cells, which leads to
the inhibition of this process.\,\,The inhibition of the process
also happens due to the enhanced level of the kinetic membrane
potential $\psi$.\,\,The parameter $\mu _0 $ is related to the lysis
and the washout of cells.

In the model, the respiratory chain of a cell is represented in two
forms: oxidized, $Q$, (12) and reduced, $q$, ones.\,\,They obey the
integral of motion $Q(t)+q(t)=L_3 $.

A change of the concentration of oxygen in the respiratory chain is
determined by Eq.~(13).

The activity of the respiratory chain is affected by the level of
NADH (14).\,\,Its high concentration leads to the enhanced endogenic
respiration in the reducing process in the respiratory chain
(parameter $k_{15} )$.\,\,The accumulation of ${\rm NADH}$ occurs as
a result of the reduction of NAD$^+$ at the transformation of
ethanol and in the Krebs cycle.\,\,These variables obey the integral
of motion NAD$^+(t)+{\rm NADH}(t)=L_2 $.

In the respiratory chain and the Krebs cycle, the substrate-linked
phosphorylation of ${\rm ADP}$ with the formation of ${\rm ATP}$
(15) is also realized.\,\,The energy consumption due to the process
${\rm ATP}\to {\rm ADP}$ induces the biosynthesis of components of
the Krebs cycle (parameter $k_{18} )$ and the growth of cells on the
substrate (parameter $k_{19} )$.\,\,For these variables, the
integral of motion ${\rm ATP}(t)+{\rm ADP}(t)=L_1 $ holds.\,\,Thus,
the level of ${\rm ATP}$ produced in the redox processes in the
respiratory chain ${\rm ADP}\to {\rm ATP}$ determines the intensity
of the Krebs cycle and the biosynthesis of proteins.

In the respiratory chain, the kinetic membrane potential $\psi $
(16) is created under the running of reducing processes $Q\to
q$.\,\,It is consumed at the substrate-linked phosphorylation ${\rm
ADP}\to {\rm ATP}$ in the respiratory chain and the Krebs
cycle.\,\,Its enhanced level inhibits the biosynthesis of proteins
and the process of reduction of the respiratory chain.

Equations (17) and (18) describe the activity of enzymes $E_1 $ and
$E_2 $, respectively.\,\,We consider their biosynthesis ($E_{1_0 } $
and $E_{2_0 })$, the inactivation in the course of the enzymatic
reaction ($n_1 $ and $n_2 )$, and all possible irreversible
inactivations ($\alpha _5 $ and $\alpha _6 )$.

Equation (19) is related to the formation of carbon dioxide.\,\,Its
removal from the solution into the environment ($\alpha _7 )$ is
taken into account.\,\,Carbon dioxide is produced in the Krebs cycle
(5).\,\,In addition, it squeezes out oxygen from the solution (13),
by decreasing the activity of the respiratory chain.

The study of solutions of the given mathematical model (1)--(19) was
performed with the help of the theory of nonlinear differential
equations [28, 29] and the methods of mathematical modeling of
biochemical systems applied and developed in
[30--47].

To solve this autonomous system of nonlinear differential equations,
we applied the Runge---Kutta--Mer\-son method.\,\,The accuracy of
solutions was set to be $10^{-8}$.\,\,To get the reliable results,
we took the duration of calculations to be $10^6$.\,\,For this time
interval, the system, being in the initial transient state,
approaches the asymptotic attractor mode, i.e., its trajectory
``sticks'' the corresponding attractor.

\section{Results of Studies }

\begin{figure*}%
\vskip1mm
\includegraphics[width=16.7cm]{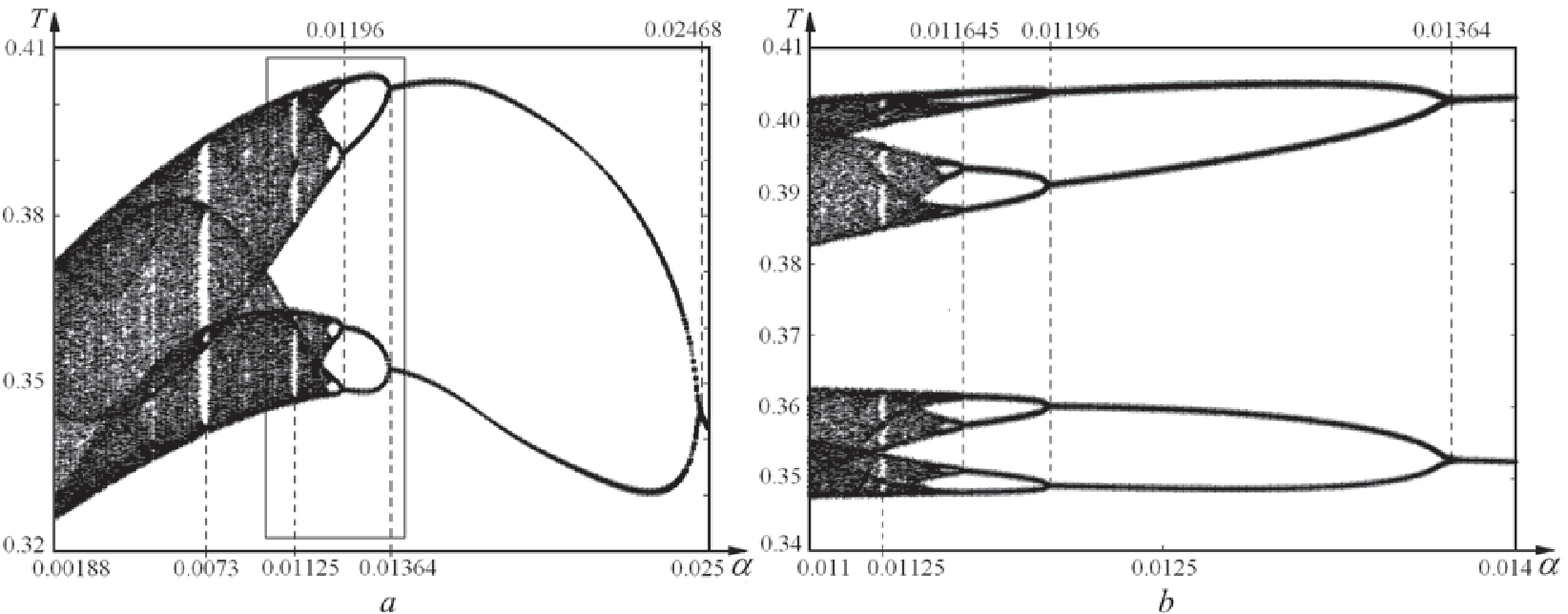}\\[2mm]
\includegraphics[width=16.7cm]{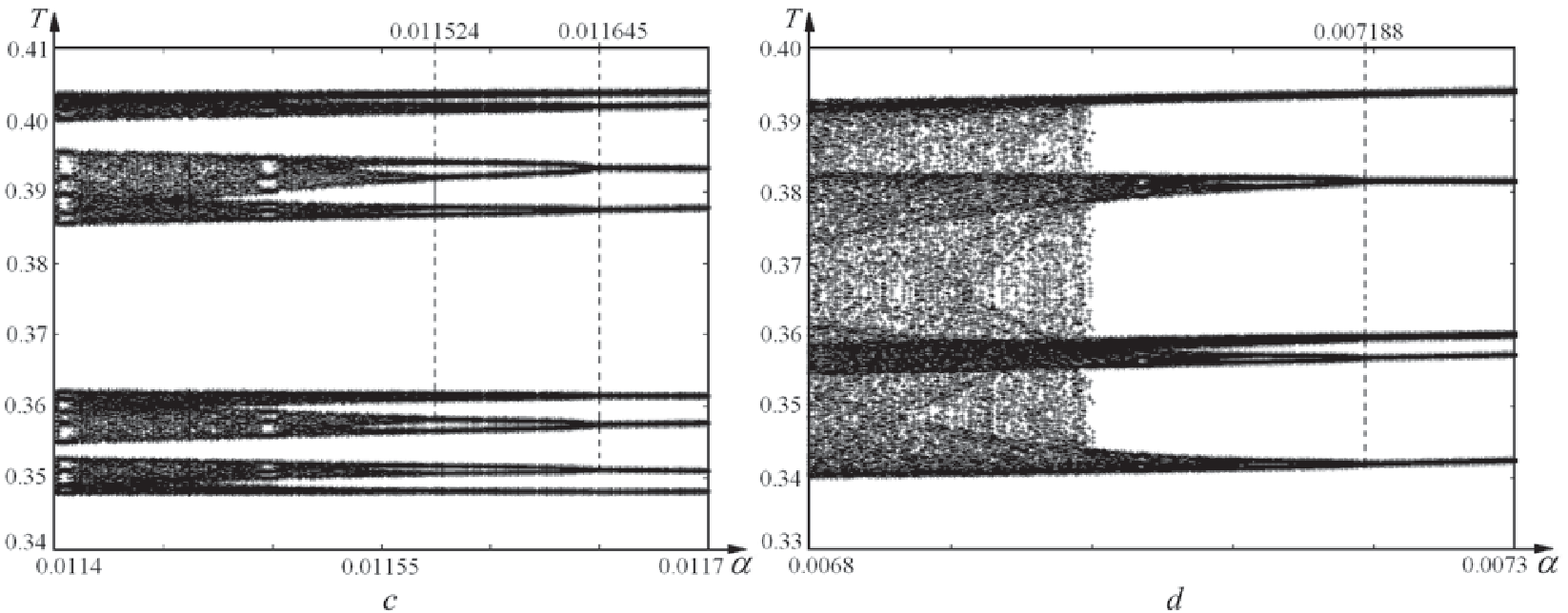}
\vskip-3mm\caption{Phase-parametric diagram for the variable $T(t)$:
\textit{a}~-- $\alpha \in (0.0018,0.025)$; \textit{b}~-- $\alpha \in
(0.011,0.014)$; \textit{c}~-- $\alpha \in (0.0114,0.0117)$;
\textit{d}~-- $\alpha \in (0.0068,0.0073)$  }\vskip1mm
\end{figure*}

With the use of the Mitchell hypothesis, we now consider the chain
of formation of the proton potential of a cell.\,\,For each turnover
of the cycle of citric acid, the specific dehydrogenases split off
four pairs of hydrogen atoms from isocitrate (5), $\alpha
$-ketoglutarate (6), succinate (7), and malate (8).\,\,Their
separation into ions H$^+$ and electrons occurs in the internal
membrane through three H$^+$-transferring loops consisting of
ubiquinone and three cytochromes.\,\,Each loop transfers two ions
H$^+$ outward the membrane, which leads to the appearance of a
transmembrane electrochemical potential (16).\,\,The acceptor of
electrons in the respiratory chain (12) is oxygen (13).\,\,Ions
H$^+$, which are accumulated on the external side of the membrane,
move again inward along the electrochemical gradient through
molecules of ${\rm F}_0 {\rm F}_1 -{\rm ATP}$ase.\,\,This transition
of ions H$^+$ from the zone with their high concentration to the
zone with a lower one is accompanied by the free energy release.
This results in the synthesis of ${\rm ATP}$ from ${\rm ADP}$ (15)
by the reaction of oxidative phosphorylation.\,\,In other words, the
continuous turnover of ions H$^+$ through the membrane
occurs.\,\,Its driving force is the transfer of electrons along the
respiratory chain.\,\,Thus, the joint self-organization of the Krebs
cycle and the respiratory chain depends on the dynamics of formation
of the proton potential.\,\,The variation of the potential is also
affected by its dissipation in other metabolic processes in a cell,
besides the current of ions H$^+$ through the membrane.\,\,In the
present work, we will study the changes in the dynamics of the Krebs
cycle depending on the dissipation of the proton potential.

We now construct the phase-parametric diagram for the multiplicity
of autooscillations of the ${\rm ATP}$ level as a function of the
dissipation of the proton potential $\alpha $ (16) (see Fig.~2) by
the method of \mbox{cutting.}\looseness=1

In the phase space of a trajectory of the system, we place the
cutting plane for $Q=0.9$.\,\,Such choice is explained by the
symmetry of oscillations of a level of the oxidized form of the
respiratory chain relative to this point in a lot of earlier
calculated modes.\,\,When the trajectory approaches the attractor,
we observe the intersection of the plane by the trajectory in a
single direction for every given value of $\alpha $.\,\,On the
phase-parametric diagram, we indicate the value of $T(t)$.\,\,If a
multiple periodic limiting cycle arises, we observe a number of
points on the plane, which coincide in the period.\,\,If a
deterministic chaos appears, the points, where the trajectory
intersects the plane, are located chaotically.

\begin{figure*}%
\vskip1mm
\includegraphics[width=16.7cm]{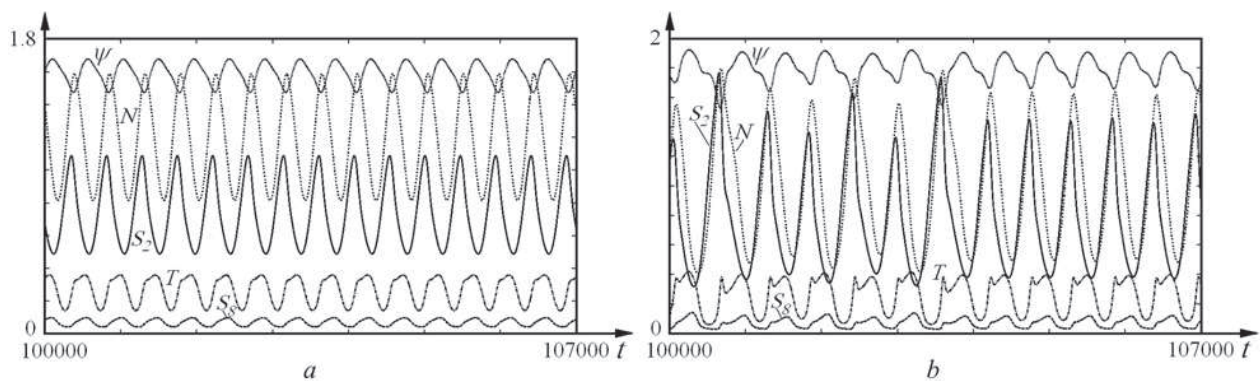}
\vskip-3mm\caption{Kinetic curves of components of the Krebs cycle
$S_2 $, $S_8 $, $N$, $\psi $, and $T$: \textit{a}~-- in the 1-fold
periodic mode $1\cdot 2^0$ for $\alpha =0.025$; \textit{b}~-- in the
chaotic mode of the strange attractor $1\cdot 2^x$ for $\alpha
=0.002$  }\vspace*{-1mm}
\end{figure*}

Let us consider the diagram in Fig.~2,~\textit{a} from right to
left.\,\,As the value of the coefficient of dissipation $\alpha $
decreases below 0.025, we see the transition from the 1-fold
periodic autooscillatory mode to the 2-fold one at the point $\alpha
^j=0.02468$.\,\,As a result of the bifurcation, the doubling of the
period of oscillations arises.\,\,For $\alpha ^{j+1}=0.01364$, we
observe the repeated doubling of the period.\,\,Then, for $\alpha
^{j+2}=0.01196$, the period of autooscillations is doubled once
more.

Let us separate a small section of the diagram for $\alpha \in
(0.011,0.014)$ (Fig.~2,~\textit{a}) and represent it in a magnified
form (Fig.~2,~\textit{b}).\,\,For $\alpha ^{j+3}=0.011645$, we see
the next bifurcation with the doubling of the period, and the
diagram becomes similar to the previous one.\,\,The subsequent
decrease in the scale of the diagram in Fig.~2,~\textit{c} reveals
the next bifurcation with the doubling of the period for $\alpha
^{j+4}=0.011524$, and the self-similarity of the diagram is
repeated.\,\,This indicates the fractal nature of the obtained
cascade of bifurcations.\,\,After the critical value of the
parameter $\alpha $ determined by the accuracy of computer-based
calculations, the deterministic chaos takes place.\,\,This means
that any appeared fluctuation under given instable modes of the real
physical system can induce a chaotic mode.\,\,This scenario of the
transition to a chaos corresponds to the Feigenbaum scenario [48].
We calculated the value of universal Feigenbaum constant by the data
on bifurcations and found that it differs from the classical one.
This means that the dynamics of system (1)--(19) cannot be reduced
completely to a one-dimensional Feigenbaum
\mbox{mapping.}\looseness=1

For $\alpha  = 0.0073$ and $\alpha =0.01125$
(Fig.~2,~\textit{a},~\textit{b}), we see the appearance of the
windows of periodi\-ci\-ty.\,\,The deterministic chaos is broken,
and the periodic and quasiperiodic modes appear.\,\,Analogous
windows of periodicity are also observed on bifurcation diagrams on
a less scale (see Fig.~2,~\textit{c}).\,\,As the coefficient of
dissipation $\alpha $ continues to decrease, the bifurcations arise
in the windows of periodicity, and the chaotic modes are seen again
(see Fig.~2,~\textit{d}).\,\,The self-similarity of the formation of
the windows of periodicity on large and small scales indicates once
more the fractality of the bifurcation diagram.

In Fig.~3,~\textit{a},~\textit{b}, we show the kinetic curves for
some components of the metabolic process of cell respiration: in the
1-fold periodic ($\alpha  = 0.025$) and chaotic ($\alpha  = 0.002$)
modes.

In Fig.~3,~\textit{a}, we see the harmonic interconnection of the
autooscillations of acetyl-CoA ($S_2 )$, which is supplied to the
cycle of citric acid, and oxalacetate ($S_8 )$ closing the
cycle.\,\,Oscillations of the level of NAD\,$\cdot $\,H\,($N$),
which transfers electrons to ubiquinone and cytochromes, occur with
the same fre\-qu\-en\-cy.\,\,These hydrogen-transferring and
electron-transferring proteins alternate in a respiratory chain, by
forming ``three loops'' in it.\,\,Electrons are transferred to
oxygen (acceptor of electrons), and ions H$^+$ move to the external
side of the membrane, by producing the gradient of the potential
$\psi $.\,\,This gradient creates the driving force for the return
of H$^+$ inward the membrane through a complex ${\rm
ATP}$-synthetase sys\-tem.\,\,This results in the creation of new
covalent bonds, through which the terminal phosphate groups join
${\rm ADP}$ with the formation of ${\rm ATP}$\,$(T)$.

\begin{figure*}%
\vskip1mm
\includegraphics[width=16.7cm]{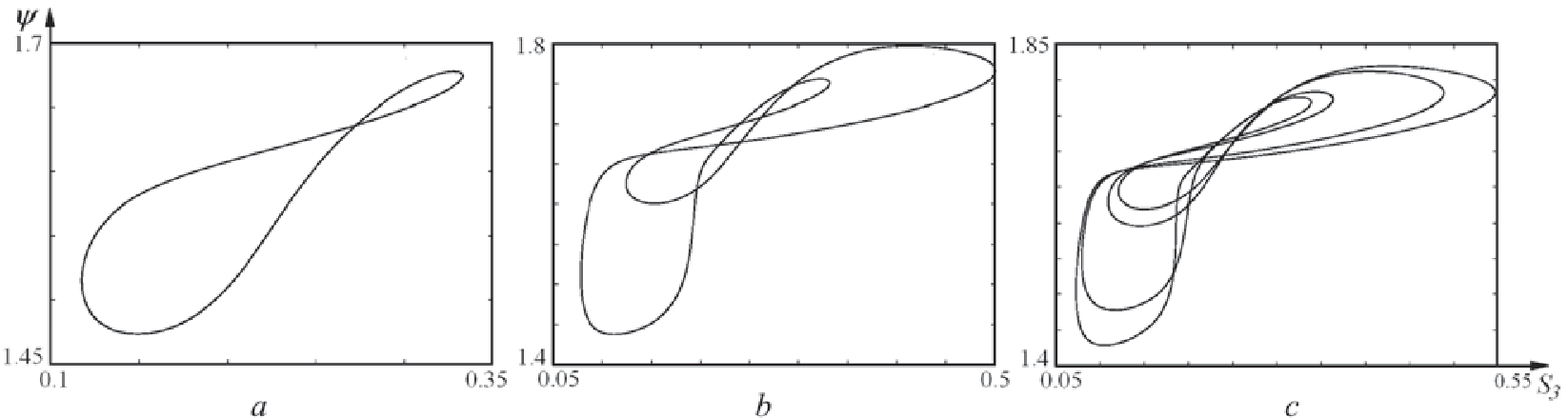}\\[2mm]
\includegraphics[width=16.7cm]{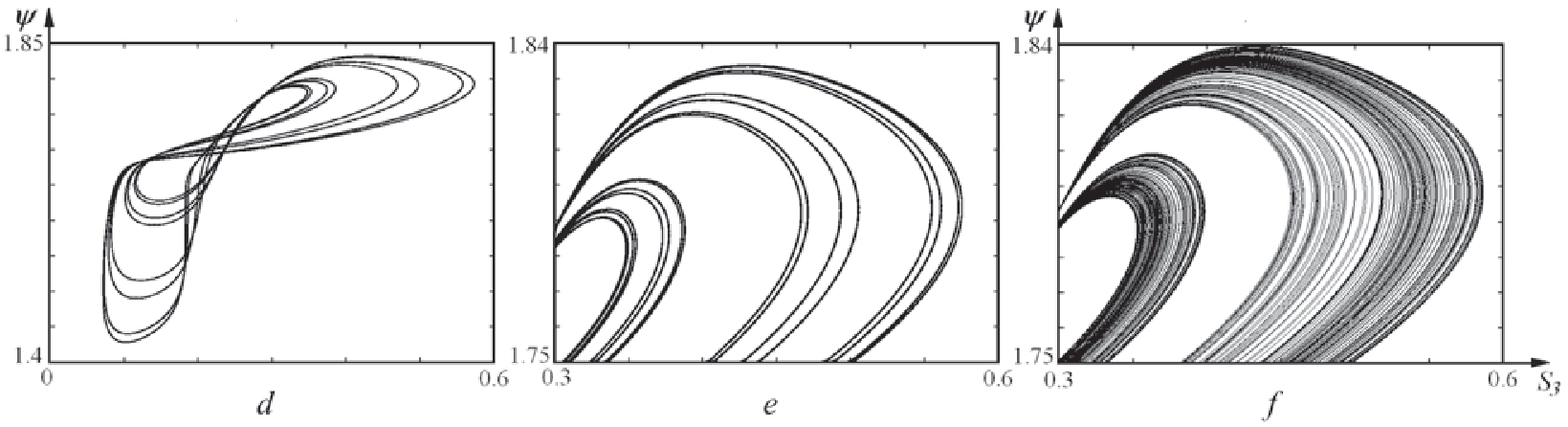}
\vskip-3mm\caption{Projections of system's phase portraits:
\textit{a}~-- regular attractor $1\cdot 2^0$, $\alpha =0.025$;
\textit{b}~-- regular attractor $1\cdot 2^1$, $\alpha =0.015$;
\textit{c}~-- regular attractor $1\cdot 2^2$, $\alpha =0.013$;
\textit{d}~-- regular attractor $1\cdot 2^3$, $\alpha =0.0117$;
\textit{e}~-- regular attractor $1\cdot 2^4$, $\alpha =0.01158$;
\textit{f}~-- strange attractor $1\cdot 2^x$, $\alpha =0.011$
}
\end{figure*}

Autooscillations arisen in the given metabolic process are regulated
by the level of dissipation of the proton potential $\alpha
$.\,\,Its decrease leads to the successive doubling of the period of
autooscillations and, as follows from calculations, to the
appearance of chaotic oscillations (Fig.~3,~\textit{b}).\,\,The
decrease in $\alpha $ means, in particular, a decrease in the
current of ions H$^+$ from the external side of the membrane to the
internal one.\,\,The time coordination between the tricarboxylic
acid cycle and the transfer of electrons and ions H$^+$ along the
respiratory chain is violated, and the process of oxidative
phosphorylation is decelerated.\,\,The deceleration of the process
of production of ${\rm ATP}$ increases the rate of the metabolic
process involving tricarboxylic acids (7).\,\,Moreover, the
frequency of the given cycle increases, which causes an increase in
the multiplicity of the period of autooscillations.\,\,As the
parameter $\alpha $ becomes critical, all given metabolic processes
become desynchronized, which leads to a chaotic mode
(Fig.~3,~\textit{b}).

The given sequence demonstrating a growth of the multiplicity of
oscillations can be observed in Fig.~4.\,\,There, we show the
sequential appearance of bifurcations and a complication of the
projections of the phase portraits of regular attractors, as the
coefficient of dissipation $\alpha $ decreases, until a strange
attractor eventually arises (see Fig.~4,~\textit{f}).\,\,Such
scenario can be explained by the existence of positive feedbacks in
the given system, which stabilize or intensify the given metabolic
processes.\,\,For the optimum value of $\alpha =0.025$, we observe
the coordination of the Krebs cycle and the rate of transfer of
charges in the respiratory chain.\,\,A decrease in $\alpha $ means
the deceleration of some metabolic processes related to the
dissipation of a membrane potential.\,\,The enhanced level of $\psi
$ blocks the respiratory chain (12)--(13), by holding it in the
reduced state.\,\,Under the new conditions, the system reveals the
self-organization, by coordinating the dynamics of the tricarboxylic
acid cycle with the transfer of electrons along the respiratory
chain.\,\,In the correspondence with the new appeared cycle, the
kinetics of variations in the proton potential gradient and the
${\rm ATP}$ level are formed.

\begin{figure*}%
\vskip1mm
\includegraphics[width=16.7cm]{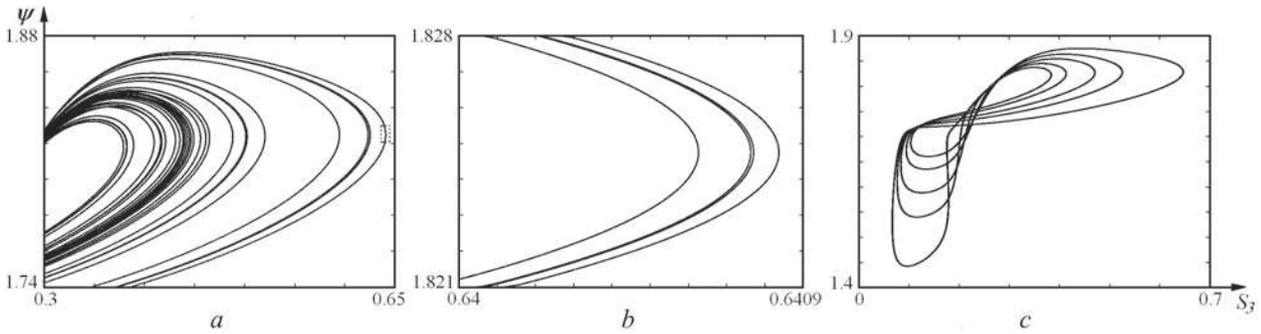}
\vskip-3mm\caption{Projections of system's phase portraits:
\textit{a}~-- strange attractor $1\cdot 2^x$, $\alpha =0.0078$,
$t\in (100000,115000)$; \textit{b}~-- strange attractor $1\cdot
2^x$, $\alpha =0.0078 \quad t\in (100000,320000)$; \textit{c}~--
regular attractor $5\cdot 2^0$, $\alpha =0.0073$  }
\end{figure*}

\begin{figure*}%
\vskip2mm
\includegraphics[width=16.7cm]{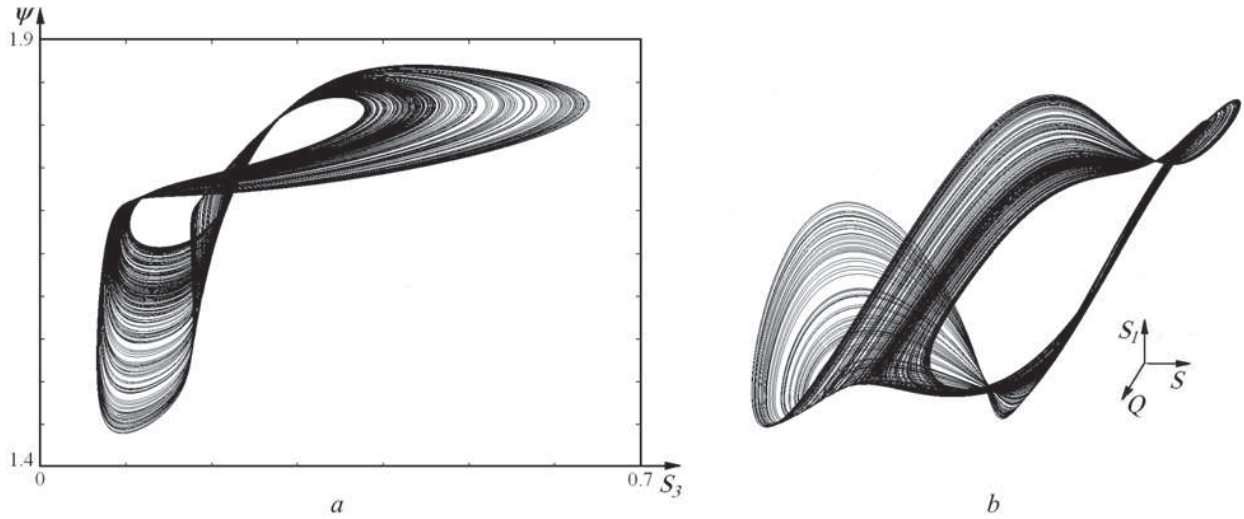}
\vskip-3mm\caption{Projections of the phase portraits of the strange
attractor $1\cdot 2^x$, $\alpha =0.0078$: \textit{a}~--
2-dimensional projection in the coordinates $(S_3,\psi )$;
\textit{b}~-- 3-dimensional projection in the coordinates $(S,S_1
,Q)$  }
\end{figure*}

We now give the example of a possible test of strange attractors for
the fractality.\,\,Let us consider the strange attractor $1\cdot
2^x$ (Fig.~5,~\textit{a}) formed for $\alpha =0.0078$.\,\,We
separate a small rectangular area of the projection of the phase
space $t\in (100000,115000)$ with a single phase curve and represent
it in Fig.~5,~\textit{b}. The calculation of a phase portrait was
executed in the interval $t\in (100000,320000)$.\,\,As is seen, the
geometric structure of the given strange attractor is repeated on
small and large scales of the projection of the phase
portrait.\,\,Each appeared curve of the projection of the chaotic
attractor is a source of the formation of new curves.\,\,Moreover,
the geometric regularity of the construction of trajectories in the
phase space is repeated.\,\,This fact confirms once more that the
phase-parametric diagrams are similar on small and large scales,
which testifies to the fractal nature of the given strange
attractor.

In Fig.~5,~\textit{c}, we present a projection of the phase portrait
of the regular attractor $5\cdot 2^0$ formed in a window of
periodicity (Fig.~2) for $\alpha =0.0073$.\,\,The deterministic
chaos is destroyed, and the periodic mode is established.\,\,The
identical windows of periodicity are observed also on smaller scales
of the diagram.\,\,Outside these windows, the chaotic modes are
formed.

\begin{figure*}%
\vskip1mm
\includegraphics[width=16.7cm]{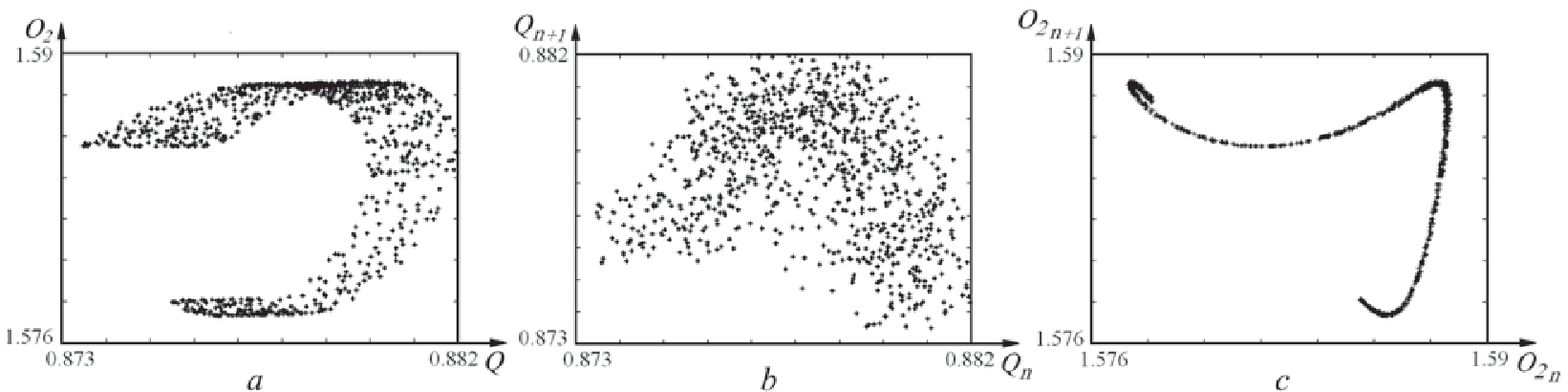}
\vskip-3mm\caption{Projection of the section by the plane $N=1.128$
(\textit{a}) and Poincar\'{e} maps (\textit{b}, \textit{c}) of the
strange attractor $1\cdot 2^x$ for $\alpha =0.0078$  }
\end{figure*}

As an example, Fig.~6,~\textit{a},~\textit{b} shows 2- and
3-dimensional projections of the phase portrait of the strange
attractor for $\alpha =0.0078$.

In Fig.~7,~\textit{a},~\textit{b},~\textit{c}, we give the
constructed projection of a section by the plane $N=1.128$ and
Poincar\'{e} maps for the given strange attractor.\,\,The choice of
a cutting surface was such that the phase trajectory $N(t)$ under a
decrease in the given component intersects it the maximally possible
number of times, and the tangency is
excluded.\,\,Figure~7,~\textit{a} indicates the chaoticity of the
given strange attractor in the plane $(Q,O_2 )$.\,\,The Poincar\'{e}
map for ($Q_n,Q_{n+1} )$ shows the instability of the phase curve
for the given component.\,\,Points of the map are located randomly
on a large part of the area.\,\,At the same time, the Poincar\'{e}
map for ($O_{2_n },O_{2_{n+1} } )$ has a quasistrip form.\,\,The
shape of the given curve is independent of the number of points of
the mapping.\,\,All points of the mapping lie on this curve.\,\,The
chaos for the given component exists only in the limits of this
curve.\,\,Along this direction of the phase space, the trajectory of
the strange attractor is stable, but \mbox{aperiodic.}\looseness=1

The chaoticity for each component has the own regularity.\,\,With
the help of Poincar\'{e} mappings, it is possible to study the
system and to find the reason for the formation of a specific type
of the strange attractor of the system.\,\,This allows one to
investigate the structural-functional connections in the metabolic
process and the reasons, for which the appearance of chaotic modes
becomes possible.

In order to uniquely identify the type of obtained attractors and to
determine their stability, we calculated the full spectra of
Lyapunov indices and their sum $\Lambda =\sum_{j=1}^{19} {\lambda _j
} $ for the chosen points.\,\,The calculation was carried out by
Benettin's algorithm with the orthogonalization of the vectors of
perturbations by the Gram--Schmidt method [29].

As a specific feature of calculations of the given indicators, we
mention the complexity of the computer-based determination of the
perturbation vectors presented by $19\times 19$ matrices.

The algorithm of calculations of the full spectrum of Lyapunov
indices consisted in the following.\,\,Ta\-king some point on the
attractor $\overline {Y_0 } $ as the initial one, we traced the
trajectory outgoing from it and the evolution of $K$ perturbation
vectors.\,\,In our case, $K = 19$ (the number of variables of the
system).\,\,The initial equations of the system supplemented by 19
complexes of equations in variations were solved numerically.\,\,As
the initial perturbation vectors, we set the collection of vectors
$\overline {b_1^0 } $, $\overline {b _2^0}, {...}, \overline
{b_{19}^0 },$ which are mutually orthogonal and normed by one.\,\,In
some time $\Delta t$, the trajectory arrives at a point $\overline
{Y_1 } $, and the perturbation vectors become $\overline {b_1^1 } $,
$\overline {b_2^1 }, {...}, \overline {b_{19}^1 } $.\,\,Their
renormalization and orthogonalization by the Gram--Schmidt method
are performed by the following scheme:
\[
\overline {b_1^1 } =\frac{\overline {b_1 } }{\left\| {\overline {b_1
} } \right\|},
\]\vspace*{-5mm}
\[
\overline {{b}'_2 } =\overline {b_2^0 } -(\overline {b_2^0 }
,\overline {b_1^1 } )\overline {b_1^1 }, \quad \overline {b_2^1 }
=\frac{\overline {{b}'_2 } }{\left\| {\overline {{b}'_2 } }
\right\|},
\]\vspace*{-5mm}
\[
\overline {{b}'_3 } =\overline {b_3^0 } -(\overline {b_3^0 }
,\overline {b_1^1 } )\overline {b_1^1 } -(\overline {b_3^0 }
,\overline {b_2^1 } )\overline {b_2^1 }, \quad \overline {b_3^1 }
=\frac{\overline {{b}'_3 } }{\left\| {{b}'_3 } \right\|},
\]\vspace*{-5mm}
\[
\overline {{b}'_4 } =\overline {b_4^0 } -(\overline {b_4^0 }
,\overline {b_1^1 } )\overline {b_1^1 } -(\overline {b_4^0 }
,\overline {b_2^1 } )\overline {b_2^1 } -(\overline {b_4 }
,\overline {b_3^1 } )b_3^1, \quad \overline {b_4^1 }
=\frac{\overline {{b}'_4 } }{\left\| {{b}'_4 } \right\|},
\]
{...}{...}{...}{...}{...}{...}{...}{...}{...}{...}{...}{...}{...}{...}{...}{...}{...}{...}{...}{...}{...}{...}{...}{...}{...}{...}{...}
\[
\overline {{b}'_{19} } =\overline {b_{19}^0 } -(\overline {b_{19}^0
},\overline {b_1^1 } )\overline {b_1^1 } -(\overline {b_{19}^0 }
,\overline {b_2^1 } )\overline {b_2^1 } -(\overline {b_{19} }
,\overline {b_3^1 } )b_3^1 -...-
\]\vspace*{-5mm}
\[
-(\overline {b_{19} },\overline {b_{18}^1 } )b_{18}^1, \quad
\overline {b_{19}^1 } =\frac{\overline {{b}'_{19} } }{\left\|
{{b}'_{19} } \right\|},
\]
Then the calculations are continued, by starting from the point
$\overline {Y_1 } $ and perturbation vectors $\overline {b_1^1 } $,
$\overline {b_2^1 }, {...}, \overline {b_{19}^1 } $. After the next
time interval $\Delta t$, a new collection of perturbation vectors
$\overline {b_1^2 } $, $\overline {b_2^2 }, {...}, \overline
{b_{19}^2 } $ is formed and undergoes again the orthogonalization
and the renormalization by the above-indicated scheme.\,\,The
described sequence of manipulations is repeated a sufficiently large
number of times, $M$.\,\,In this case in the course of calculations,
we evaluated the sums
\[
Z_1 =\sum\limits_{i=1}^M {\ln \left\| b_{1}^{\prime i}  \right\|},
\]
\[
 Z_2
=\sum\limits_{i=1}^M {\ln \left\| b_2^{\prime  } \right\|},
\]
 ........................,\vspace*{-3mm}
\[
Z_{19} =\sum\limits_{i=1}^M {\ln \left\| b_{19}^{\prime i} \right\|}
\]
which involve the perturbation vectors prior to the renormalization,
but after the normalization.

The estimation of 19 Lyapunov indices was carried out in the
following way:
\[
\lambda _j =\frac{Z_j }{MT}, \quad j=1, 2, ..., 19.
\]
Below for the sake of comparison, we present the spectra of Lyapunov
indices for some modes of the system. For brevity without any loss
of information, we give the values of indices up to the fourth
decimal point.

The ratios of the values of Lyapunov indices $\lambda _1 >$
$>\lambda _2
>\lambda _3 >\,...\,>\lambda _{19} $ serve as the criterion of the validity of
calculations.\,\,For a regular attractor, we have obligatorily
$\lambda _1 -\lambda _{19} $.\,\,The remaining indices can be also
$\approx 0$ in some cases.\,\,In some other cases, they are
negative.\,\,The zero value of the first Lyapunov index testifies to
the presence of a stable limiting cycle.

For a strange attractor, at least one Lyapunov index must be
positive.\,\,After it, the zero index follows.\,\,The next indices
are negative.\,\,The presence of negative indices means the
contraction of system's phase space in the corresponding directions,
whereas the positive indices indicate the dispersion of
trajectories.\,\,Therefore, there occurs the mixing of trajectories
in narrow places of the phase space of the system, i.e., there
appears the deterministic chaos.\,\,The Lyapunov indices contain
obligatorily the zero index, which means the conservation of the
aperiodic trajectory of an attractor in some region of the phase
space and the existence of a strange attractor.

For $\alpha  = 0.025$, the regular attractor $1\cdot 2^0$
arises.\,\,We have $\lambda _1 -\lambda _{19} $: .0000, --.0001,
--.0002, --.0040, --.0147, --.0230, --.0241, --.0254, --.0355,
--.0355, --.0355, --.0355, --.0362, --.0369, --.0897, --.1091,
--.1090, --.1221, --.1507. $\Lambda $ = --.8872.

For $\alpha  = 0.015$~-- regular attractor $1\cdot 2^1$; $\lambda _1
-\lambda _{19} $: .0000, --.0002, --.0006, --.0040, --.0135,
--.0193, --.0232, --.0284, --.0316, --.0322 --.0322, --.0322,
--.0406, --.0436, --.0849, --.1000, --.1000, --.1192, --.1530.
$\Lambda $ = --.8586.

For $\alpha  = 0.013$~-- regular attractor $1\cdot 2^2$; $\lambda _1
-\lambda _{19} $: .0000, --.0002, --.0008, --.0040, --.0127,
--.0194, --.0237, --.0297, --.0304, --.0320, --.0320, --.0320,
--.0409, --.0432, --.0845, --.0981, --.0981, --.1183, --.1531.
$\Lambda $ = --.8531.

For $\alpha  = 0.0117$~-- regular attractor $1\cdot 2^3$; $\lambda
_1 -\lambda _{19} $: .0000, --.0001, --.0002, --.0040, --.0130,
--.0194, --.0232, --.0295, --.0306, --.0322, --.0322, --.0322,
--.0406, --.0448, --.0834, --.0969, --.0969, --.1179, --.1533.
$\Lambda $ = --.8503.

For $\alpha  = 0.01158$~-- regular attractor $1\cdot 2^4$; $\lambda
_1 -\lambda _{19} $: .0000, .0000, --.0002, --.0040, --.0130,
--.0194, --.0231, --.0297, --.0305, --.0322, --.0322, --.0322,
--.0406, --.0443, --.0840, --.0968, --.0967, --.1178, --.1532.
$\Lambda $ = --.8500.

For $\alpha  = 0.01153$~-- strange attractor $1\cdot 2^x$; $\lambda
_1 -\lambda _{19} $: .0001, .0000, --.0002, --.0040, --.0131,
--.0194, --.0231, --.0298, --.0305, --.0322, --.0322, --.0322,
--.0407, --.0443, --.0840, --.0967, --.0967, --.1178, --.1530.
$\Lambda $ = --.8496.

For $\alpha  = 0.011$~-- strange attractor $1\cdot 2^x$; $\lambda _1
-\lambda _{19} $: .0003, .0000, --.0002, --.0040, --.0132, --.0193,
--.0228, --.0305, --.0306, --.0317, --.0317, --.0317, --.0404,
--.0443, --.0842, --.0962, --.0961, --.1176, --.1529. $\Lambda $ =
--.8471.

For $\alpha  = 0.0078$~-- strange attractor $1\cdot 2^x$; $\lambda
_1 -\lambda _{19} $: .0007, .0000, --.0002, --.0040, --.0129,
--.0200, --.0213, --.0297, --.0311, --.0326, --.0326, --.0326,
--.0415, --.0449, --.0831, --.0928, --.0928, --.1164, --.1527.
$\Lambda $ = --.8401.

For $\alpha $ = 0.00735260, we have the transition between the
strange attractors $1\cdot 2^x\leftrightarrow 5\cdot 2^x$.\,\,$
\lambda _1 -\lambda _{19} $:  .0003,  .0000,  --.0002, --.0040,
--.0125, --.0200,  --.0210,  --.0296,  --.0309,  --.0322, --.0322,
--.0322, --.0406, --.0459, --.0829, --.0926, --.0926, --.1168,
--.1532. $\Lambda $ = --.8390.

For $\alpha  = 0.00735255$~-- strange attractor $5\cdot 2^x$;
$\lambda _1 -\lambda _{19} $: .0003, .0000, --.0002, --.0040,
--.0125, --.0200, --.0210, --.0296, --.0310, --.0321, --.0321,
--.0321, --.0405, --.0459, --.0829, --.0926, --.0926, --.1170,
--.1534. $\Lambda $ = --.8392.

For $\alpha  = 0.0073$~-- regular attractor $5\cdot 2^0$; $\lambda
_1 -\lambda _{19} $: .0000, --.0002, --.0008, --.0039, --.0118,
--.0200, --.0209, --.0296, --.0308, --.0321, --.0321, --.0321,
--.0405, --.0460, --.0828, --.0926, --.0926, --.1171, --.1533.
$\Lambda $ = --.8390.

For $\alpha  = 0.00715$~-- regular attractor $5\cdot 2^1$; $\lambda
_1 -\lambda _{19} $: .0000, --.0001, --.0004, --.0047, --.0113,
--.0201, --.0209, --.0297, --.0307, --.0322, --.0322, --.0322,
--.0405, --.0460, --.0827, --.0925, --.0924, --.1171, --.1533.
$\Lambda $ = --.8389.

For $\alpha  = 0.0071$~-- regular attractor $5\cdot 2^2$; $\lambda
_1 -\lambda _{19} $: .0000, --.0001 --.0002, --.0040, --.0123,
--.0201, --.0208, --.0297, --.0307, --.0322, --.0322, --.0322,
--.0405, --.0461, --.0826, --.0924, --.0924, --.1171, --.1533.
$\Lambda $ = --.8388.

For $\alpha  = 0.0070$~-- strange attractor $5\cdot 2^x$; $\lambda
_1 -\lambda _{19} $: .0004, .0000, --.0002, --.0041, --.0126,
--.0200, --.0208, --.0297, --.0306, --.0325, --.0325, --.0325,
--.0406, --.0462, --.0826, --.0922, --.0922, --.1168, --.1533.
$\Lambda $ = --.8388.

For $\alpha  = 0.00695$, we have the transition between the strange
attractors $5\cdot 2^x\leftrightarrow 1\cdot 2^x$. $\lambda _1
--\lambda _{19} $:  .0006,  .0000,  --.0002, --.0040, --.0127,
--.0199,  --.0209,  --.0298,  --.0306,  --.0326, --.0326, --.0326,
--.0407, --.0462, --.0827, --.0921, --.0920, --.1165, --.1530.
$\Lambda $ = --.8385.

For $\alpha  = 0.002$~-- strange attractor $1\cdot 2^x$; $\lambda _1
-\lambda _{19} $: .0007, .0000, --.0002, --.0040, --.0122, --.0190,
--.0205, --.0290, --.0308, --.0324, --.0324, --.0324, --.0410,
--.0477, --.0812, --.0871, --.0870, --.1167, --.1528. $\Lambda $ =
--.8256.

These calculations allow us to uniquely identify the type of
attractors in the corresponding modes.\,\,A decrease in the
coefficient of dissipation of the transmembrane potential causes
sequential bifurcations in the 1-fold periodic mode $1\cdot 2^0\to
1\cdot 2^n$, until the modes of the strange attractors $1\cdot 2^x$
arise eventually.\,\,We observe the ``order-chaos''
transition.\,\,As the dissipation decreases further, the windows of
periodicity appear.\,\,In them, the 5-fold periodic modes $5\cdot
2^0\to 5\cdot 2^n$ are formed.\,\,As a result of sequential
bifurcations, the strange attractors $5\cdot 2^x$ arise.\,\,We
observe also the ``chaos-chaos'' transitions: $1\cdot
2^x\leftrightarrow 5\cdot 2^x$ and $5\cdot 2^x\leftrightarrow 1\cdot
2^x$.\,\,Thus, the metabolic system has possibilities for the
self-organization and the adaptation to varying
con\-ditions.\looseness=1

By the given Lyapunov indices for strange attractors, we determine
the KS-entropy (the Kolmogorov--Sinai entropy) [49].\,\,By the Pesin
theorem [50], the KS-entropy $h $corresponds to the sum of all
positive Lyapunov characteristic indices.

The KS-entropy allows us to judge the rate, with which the
information about the initial state of the system is lost.\,\,The
positivity of the given entropy is a criterion of the chaos.\,\,This
gives possibility to qualitatively estimate the properties of
attractor's local stability.

We determine also the quantity inverse to the KS-entropy, $t_{\min }
.$ This is the time of a mixing in the system.\,\,It characterizes
the rate, with which the initial conditions will be
forgotten.\,\,For $t\ll t_{\min } $, the behavior of the system can
be predicted with sufficient accuracy.\,\,For $t>t_{\min } $, only a
probabilistic description is possible.\,\,The chaotic mode is not
predictable due to the loss of the memory of initial
conditions.\,\,The quantity $t_{\min } $ is called the Lyapunov
index and characterizes the ``horizon of predictability'' of a
strange attractor.

In order to classify the geometric structure of strange attractors,
we calculated the dimension of their fractality.\,\,The strange
attractors are fractal sets and have the fractional
Hausdorff--Besicovitch dimension.\,\,But its direct calculation is a
very labor-consuming task possessing no standard algorithm.
Therefore, as a quantitative measure of the fractality, we
calculated the Lyapunov dimension of attractors by the Kaplan--Yorke
formula [51, 52]:
\[
D_{F_r } =m+\frac{\sum\limits_{j=1}^m {\lambda _j } }{\left|
{\lambda _{m+1} } \right|},
\]
where $m$ is the number of the first Lyapunov indices ordered by
their decreasing.\,\,Their sum $\sum_{j=1}^m {\lambda _j } \geqslant
0$, and $m+1$ is the number of the first Lyapunov index, whose value
$\lambda _{m+1} <0$.

For the above-considered strange attractors, we obtained the
following indices.

$1\cdot 2^x$ ($\alpha  = 0.01153$): $h = 0.0001$, $t_{\min }  =
10000$, $D_{F_r } = 2.5$;

$1\cdot 2^x$ ($\alpha  = 0.011$): $h = 0.0003$, $t_{\min }  =
3333.3$, $D_{F_r } = 3.5$;

$1\cdot 2^x$ ($\alpha  = 0.078$): $h = 0.0007$, $t_{\min }  =
1428.6$, $D_{F_r } = 5.5$;

$1\cdot 2^x\leftrightarrow 5\cdot 2^x$ ($\alpha  = 0.00735260$): $h
= 0.0003$, $t_{\min }  = 3333.3$, $D_{F_r } = 3.5$;

$5\cdot 2^x$ ($\alpha  = 0.00735255$): $h = 0.0003$, $t_{\min }  =
3333.3$, $D_{F_r } = 3.5$;

$5\cdot 2^x$ ($\alpha  = 0.0070$): $h = 0.0004$, $t_{\min }  =
2500$, $D_{F_r } = 5.0$;

$5\cdot 2^x\leftrightarrow 1\cdot 2^x$ ($\alpha  = 0.00695$): $h =
0.0006$, $t_{\min }  = 1666.7$, $D_{F_r } = 6$.

$1\cdot 2^x$ ($\alpha  = 0.002$): $h = 0.0007$, $t_{\min }  =
1428.6$, $D_{F_r } = 5.5$.

By the values of calculated indicators, we may judge the distinction
of the structures of the given strange attractors.\,\,The higher the
Lyapunov dimensionality, the more pronounced the chaoticity of an
attractor in the phase space.\,\,As the Lyapunov dimensionality
decreases, the phase curves approach one another in an element of
the phase space (cf.\,\,the Lyapunov dimensionalities and the
chaoticity of strange attractors in Fig.~4,~\textit{f} and
Fig.~6,~\textit{a}).\,\,Starting from the structure of a strange
attractor, we can conclude about the degree of stability of various
modes and about the adaptation of the metabolic process in a cell to
external influences.

\section{Conclusions}

With the help of the mathematical model, we have studied the
self-organization of the Krebs cycle and the respiratory chain of a
cell depending on the dissipation of the transmembrane potential.
The work is based on the Mitchell hypothesis about the formation of
a transmembrane potential and its dissipation under phosphorylation
and its consumption in other metabolic processes.\,\,The bifurcation
diagram is constructed, and a scenario of variation of the
multiplicity of the autooscillatory metabolic process is
found.\,\,As the dissipation decreases, the multiplicity of the
cycle is doubled by the Feigenbaum scenario, until the aperiodic
modes of strange attractors arise eventually.\,\,They are
transformed in stable periodic modes as a result of the
self-organization.\,\,This means that the metabolic process is
adapted to varying conditions.\,\,The Poincar\'{e} sections and maps
for attractors are constructed.\,\,The full spectra of Lyapunov
indices and the divergences for various modes are calculated.\,\,For
some strange attractors, we have calculated the KS-entropies,
``horizons of predictability,'' and the Lyapunov dimensionalities of
attractors.\,\,The obtained results allow one to study the
structural-functional connections, by which the transmembrane
potential affects the dynamics of the Krebs cycle and the
respiratory chain, and to study the physical laws of a
self-organization in the metabolic process in cells.

\vskip3mm \textit{The work is supported by the project
No.\,0113U001093 of the National Academy of Sciences of Ukraine.}

\vspace*{-5mm}
\rezume{%
В.Й.\,Грицай}{ПОКАЗНИКИ ЛЯПУНОВА\\ ТА ВІДОБРАЖЕННЯ ПУАНКАРЕ\\ В
ДОСЛІДЖЕННІ СТІЙКОСТІ ЦИКЛУ КРЕБСА} {В даній роботі за допомогою
математичної моделі продовжується дослідження метаболічного процесу
циклу Кребса. Вперше досліджується узгодженість та стійкість циклу
трикарбонових кислот в залежності від дисипації трансмембранного
потенціалу, утвореного дихальним ланцюгом в плазматичній мембрані
клітини. Побудована фазопараметрична характеристика залежності
динаміки зміни рівня АТФ від величини даного параметра. Знайдено
сценарій формування багатократних автоперіодичних та хаотичних
режимів. Побудовано Пуанкаре перерізи та відображення. Досліджено
стійкість режимів та фрактальність знайдених біфуркацій. Розраховані
повні спектри показників Ляпунова, дивергенції, КС-ентропії,
горизонти передбачуваності та ляпуновські розмірності дивних
атракторів. Зроблено висновки про структурно-функціональні зв'язки,
що визначають залежність циклічності дихання клітини від
синхронізації функціонування циклу трикарбонових кислот та
електротранспортного ланцюга.}



\end{document}